# Trust, Professional Vision and Diagnostic Work

Mark Rouncefield
Department of Computing, Lancaster University
m.rouncefield@lancaster.ac.uk

Rob Procter
Department of Computer Science, Warwick University and Alan Turing Institute for Data Science and AI
rob.procter@warwick.ac.uk

Peter Tolmie
Wirtschaftsinformatik und Neue Medien, Universität Siegen
Peter.Tolmie@uni-siegen.de

Abstract. In this paper we consider some empirical materials from our ongoing research into forms of everyday detection and diagnosis work in healthcare settings, and how these relate to issues of trust; trust in people, in technology, processes and in data.

"One of the basic conditions of any constitutive practice is a mutual commitment to rules of engagement in that practice – that is, all parties to the interaction must understand that they are engaged in the same practice, must be competent to perform the practice, must actually perform competently and assume this also of the others." (Watson 2009)

## Introduction

This workshop paper considers some empirical materials from our ongoing research into forms of everyday detection and diagnosis work in healthcare settings, and how these relate to issues of trust; trust in people, in technology, processes and in data. The raw and as yet relatively under-analysed empirical data points to some of the features of everyday work in the Pathology Lab, as well as some design issues associated with developing AI systems intended to support 'trusted' processes of detection and diagnosis. The setting for this particular research, and the collection of these materials, was a number of pathology labs where the pathologists were looking for various forms of cancer and were interviewed about developing working practices and the impact of the movement towards digitalisation of images and their thoughts about the introduction of AI tools to assist them. Pathology lab work involves the employment of particular and complex perceptual skills to find what may seem small features or irregularities in the complex visual environment of a digitized image, and interpretative skills to classify them appropriately, in particular, as being 'suspicious' 'cancerous' and thereby worthy of further diagnostic investigation. Whilst such work is clearly 'routine' for practitioners, as Giddens (1984) reminds us, such routine work requires detailed analysis of its accomplishment:

"It is a major error to suppose that these phenomena need no explanation, that they are simply repetitive forms of behaviour carried out 'mindlessly'. On the contrary, as Goffman (together with ethnomethodology) has helped to demonstrate, the routinized character of most social activity is something that has to be 'worked at' continually by those who sustain it in their day-to-day conduct." (Giddens1984: 86).

The accomplishment we are especially interested in is the accomplishment of trust in the process of everyday work – not least if we are intending to design, build or evaluate diagnostic technologies intended to support the work of professionals.

# Everyday Work in the Pathology Lab

As with many other forms of diagnostic work the practice of 'reading' and interpreting these digital images, (like the original Pathology Lab practice of 'reading' glass slides through a microscope), calls for the exercise of a subtle, learned combination of reasoning, knowledge and skill. In previous work (Hartswood et al. 2000, 2002) we have considered such practices as constitutive of some form of 'professional vision' –"socially organized ways of seeing and understanding events that are answerable to the distinctive interests of a particular social group" (Goodwin 1994: 606). As we have observed in other diagnostic settings, diagnostic work in the Pathology Lab involves, requires even, sets of tried and tested repertoires of 'manipulations' that are an integral part of the embodied practice of uncovering or realizing phenomenon in the images – of revealing cancerous cells. These manipulations and the accompanying diagnosis are examples of 'midenic' reasoning, that is reasoning, assessing, evaluating, diagnosing, making judgements, whilst engaged in the actual flow of activity. Here is one pathologist talking about some of the ways in which the digitised images are manipulated in order to develop trust in what is being seen:

*P1: Once you're looking at the image is the same as what's down the microscope training and how to manipulate the images and the settings and things so that it looks as similar as possible to the glass.*
*I: OK, so what sort of manipulations would that involve?*
*P1: I mean, so there's the very basic manipulation, such as just moving the image around the screen so that you can look at different parts of the image. And obviously zooming in and out with magnification and then an added benefit of the digital is you have measuring tools, so you can very accurately measure for example, size of a tumour or the distance of a tumour to a surgical margin. And provides accurate information.*

At the same time such activities are embedded in a set of professional expectancies, professional 'trust' and a 'professional vision' concerning how to go about everyday diagnostic work. This involves translating features made visible in the digitised image into an appropriate organisational and professional formulation – particularly in terms of possible diagnosis and possible treatment. This is what 'professional vision' looks like in action, it is concerned with the activities of the individual relative to some particular professional set of expectancies: 'The relevant unit for the analysis of the intersubjectivity at issue here is thus not these individuals as isolated entities but (. . .) a profession, a community of competent practitioners, most of whom have never met each other but nonetheless expect each other to be able to see and categorize the world in ways that are relevant to the work, tools, and artifacts that constitute their profession' (Goodwin 1994: 615). For example, in this interview extract the pathologist discusses the basic set up of the equipment – this is a professional talking about the tools of his trade, and what he regards as their 'professional' use:

*P3: … when you're looking at slide the slide or an image you're not cruising, kind of trying to find whatever you'll come through. You'll actually have, depending on what the type of the biopsy or type specimen you have, you already have a plan you have a set of questions you're answering mentally while you're going through, and it's, this is something we try to explain to the computer scientists, we're still kind of in the process. I think they're getting the idea because trying to get you know some one of the projects about artificial intelligence. So because from the clinical background, clinical training and the experience you when you look at the biopsy for example, you know these are the questions, you know, the clinical history. You know the these are the things I need to be looking for in on each specimen. These are different set of questions and before that even you look for adequacy you look for representation and so on. But it kind of getting to the nutshell there are a set of questions you kind of look for their answers, while when you are looking at the slides these are related to the type of the biopsy the clinical history and of course the presenting features on the slide itself. Because sometimes you look at the biopsy kind of things are unexpected. So your mental kind of pathway kind of changes accordingly as well.*

Diagnosis appears then as a social process of a specific 'community of practice' to which its members are accountable. Being a competent practitioner involves being able to accountably distinguish between what is 'normal' and what is 'abnormal' in a digital image or a microscope slide and understanding the range of manipulations and shared professional interactional practices that make what is 'normal' or 'abnormal' witnessable and accountable. It is, as Garfinkel et al. point out, the 'intertwining of worldly objects and embodied practices' (1981, p.165) that produces the recognizable and accountable diagnoses and decisions. For example, in this next interview extract the pathologists are talking about the processes they follow in 'looking' at an image and making their decision:

*I: How do you organize looking at an image, is there a particular way you do it?*
*P1: So, for me personally I really focus on the low power first of all. So, first of all, to make sure that I'm happy we've got a complete image and also it gives me a sense of where I might need to focus in on so rather than going up into a medium or higher power in sort of the first field, I don't do that. I tend to scan the whole image on quite a low power first to get a feel for where I might need to go in and look at on higher power. And then I might sort of rescan quite quickly on a higher power and just go into those areas where I want to focus on in. So, for a much higher power. So, like I might scan the whole thing on the equivalent of like a * 2 and then sort of start to look around on a * 4 but do that quite quickly because I know where I want to go and then quickly go into sort of * 10 * 20. That sort of thing. To really focusing on getting the information I want, as you say, to either be able to quickly dismiss that area which had caught my eye initially, or to be thinking OK, now this is something and you know these are the steps I need to take. This is a tumour or whatever.*
*P1: … there are some cases where it all looks a little bit different and you actually you just have to take your time and probably scan on a higher power. Like maybe * 10 and then you know be going up and down up and down a bit more, but particularly with excision cases where you've got quite a bit of normal tissue as well as possibly a tumour, it is easier to sort of scan on low power at and then just go in and out where you need to. There is particularly like a small biopsy you because you've got less material to actually make a diagnosis from, you're probably going to be spending longer per image than you would on a bigger case because you've got to get more information from that one image.*

In another interview:

*I: What kinds of things are you looking for, and does that depend on the particular suspected cancer?*
*P2: Yes, so, If I'm looking at a piece of tissue, the first thing I would try and decide is what the tissue represents, what normal site, or what normal anatomy I can see in that issue. So, if as often is the case, you would find a bit of residual normal tissue. I will identify that as the pancreas or lung whatever it maybe I'll try identify that and then look for piece of tissue that don't fit into what I would call his normal morphology for that particular area. So, if I'm looking at a piece of liver and I can see normal liver tissue, I'll keep looking at the rest of the tissue in that biopsy. And as soon as I find the focus that looks different from that. That's when I start looking at it closer and determining if that is neoplastic or meaning cancer or something that is totally unrelated and just in inflammation or an abscess. Things like that.*
*P2: ... if I'm looking at a piece of lung tissue, I would expect to find almost big empty spaces lined by the alveoli or the lung epithelial tissue and. If instead of that I'm finding solid areas rather than empty spaces if I'm finding all these spaces filled with either cells or any other abnormal material. Then I do know that it is abnormal. Similarly in the colon when I'm looking at a piece of intestine, I am trying to see if I can identify all the normal different layers of the colon. Now the moment I find layer or group of cells within a particular layer that looks abnormal, I would then zoom in onto their particular focus to find out what that what that is.*

Diagnosis should then be regarded as a material, collaborative process involving technologies, expert skills, and careful sensory and sensitive collaborative engagement with others. Some diagnostic activity requires what might be regarded as rational everyday knowledge, some demands specific 'scientific' epistemic practices of measurement, representations and calculations. As Heritage (1984) argues, this kind of analysis; "vividly demonstrates that where sociological research encounters institutional domains in which values, rules or maxims of conduct are overtly invoked, the identification of these latter will not provide an explanatory terminus for the investigation. Rather their identification will constitute the first step of a study directed at discovering how they are perceivedly exemplified, used, appealed to and contested." (Heritage 1984) In this interview extract we hear a skilled practitioner talking about the differences between his approach and those of the novice or trainee in terms of an appreciation of 'context':

*P1: for example, surgeons will use diathermy to burn blood vessels when they're taking out a specimen to reduce blood loss from the patient and that you get a burn artifact on that issue. Yeah, essentially sort of shrivelled and cooked and what happens to the nuclei at the edge there is they look quite distorted. And if you see that in context you know it's just the margin, it's just diathermy artifact. But if you were to see that out of context, you might be worried about the nuclei because they would look a bit darker. That sort of thing. But you see that on glass also, but would it be immediately recognizable for that to you or would be to me. If I was a junior who hadn't been doing pathology for very long, then I might be worried about that, but if you've been, if you've got the experience and you know what it relates to, then it's fine. You can just miss it. OK?*

This difference also emerges in the explanation of decisions:

*P1: ... so as we're going through our training, the trainees will describe all of these features. And the more experience you get, you sort of cut to the chase more and you won't include all of that. I will include things like that if I found it very difficult to come to a decision. If I think that there's a potential. Not that I'm wrong, but I'm giving. I'm giving an indication of how sure I am about something. So, for example, you know it might be difficult. I might have shown it to several colleagues. We might have been thinking OK, this could be positive. It might not be positive. These are the things we've taken into account and this is the*

*conclusion we've come to. But if it's a really, really straightforward case, even if I've had some of the same thought process is if I've been very, it's been very easy and quick to sort of dismiss that and get to the crux of the matter. Then I wouldn't put all of that in a report.*

This is where we anticipate a concern with 'trust' can be useful, in a way that 'professional vision' perhaps might not be, in unpacking some of the important features and characteristics of everyday rules, values and conduct. 'Professional vision' provides a gloss for activities that perhaps need to be more thoroughly understood. Notably, the term tells us little about trust and the role of trust in everyday working practice. Working collaborations with colleagues and within organisational structures obviously presupposes some form or forms of trust. It is not only individuals that must be trusted, but also, and inevitably, organisational processes and procedures, as well as different tools and data that permeate and mediate relationships and enable (or disable) trust. 'Trust' is clearly a difficult topic (though frequently treated unproblematically as a mere 'resource') and has produced various philosophical and social scientific concepts and theories (Luhman 2018, Gambetta 2000) identifying the grounds of trust in an individual's reputation, performance and appearance with a range of different relational and cultural dimensions, none of which are necessarily adequately encompassed in the notion of 'professional vision'.

Two particular ideas emerge from our empirical work. Firstly, that trust is not merely concerned with individuals, processes or technologies but also 'trusted data'. Trusted data can be an important factor in fostering trust between workers. Secondly that the 'temporality' of data should be acknowledged; understandings of trust, who can be trusted, and what constitutes a 'trustable' procedure or 'trustable' data, change over time. and is especially relevant in the continued maintenance of trust. Future work will explore these ideas further using some of Garfinkel's ideas about trust (Watson 2009), whereby trust becomes 'a phenomenon of ordinary membership', considering the expectancies of trust that precede interaction and those aspects of trust that emerge as part and parcel of the production and accomplishment of everyday professional work: "One of the basic conditions of any constitutive practice is a mutual commitment to rules of engagement in that practice – that is, all parties to the interaction must understand that they are engaged in the same practice, must be competent to perform the practice, must actually perform competently and assume this also of the others." (Watson 2009)

# Technology, Diagnosis and Trust

Our research on diagnostic work and ideas about trust is formulated and presented in the belief that a detailed understanding of everyday trust and diagnostic practices should be a precursor to the design and redesign of computer assisted detection and diagnosis technologies. We are concerned with understanding the impact such tools might have on the situated, collaborative practical activity of detection and diagnosis that we have observed. How might such tools mesh with current diagnostic practices? What future practices should be developed? In terms of providing diagnostic support, what design features of the technology might cause people to 'trust' or 'mistrust' it? We suggest that, when considering how trust might be achieved it may be useful to consider issues of 'professional trust' trust. As our pathologists suggest they are trusted to act in a professional way, and any contestation of a decision must be 'accountable' – reasons, professional reasons, have to be provided. Trust here is not a binary value, but rather it is fine grained and is an ongoing social accomplishment as part of the work and the demonstration of competence.

The work of the Pathology Lab, whilst it has its individual components, also has a profoundly social character and we might consider the possible impact of technology on these working arrangements and practices. Related to this is the wider issue of how technology should be designed and implemented in healthcare settings. As technology becomes ubiquitous in healthcare settings, and as the technology becomes wrapped up in the complexities of organisational working, so the challenges of systems design correspondingly increase since the 'design problem' becomes not merely the design and development of new healthcare technologies but the integration of IT systems with existing and developing work practices.

# Acknowledgements

This work was partly funded through the PathLake project and the Alan Turing Institute for Data Science and AI.